\documentclass[article]{rstransa}%%%%where rstrans is the template name

\usepackage[numbers]{natbib}
\usepackage{lipsum}
\usepackage{multirow}
\usepackage{ulem}
\setstcolor{red}

\titlehead{Research}

\begin{document}

%%%% Article title to be placed here
\title{Centaur 29P/Schwassmann-Wachmann 1 and its near-nucleus environment from a stellar occultation}

\author{
C.~L.~Pereira$^{1,2}$, 
F.~Braga-Ribas$^{3,2}$, 
B.~Sicardy$^{4}$, % group leader
B.~E.~Morgado$^{5,2}$, % código para ajuste duplo (fora do SORA) + contribuição da correção das imagens
J.~L.~Ortiz$^{6}$, % group leader
M.~Assafin$^{5,2}$ % PRAIA e discussões sobre a remoção da coma
R.~Miles$^{7}$, % Astrometric positions
J.~Desmars$^{8,9}$, % prediction
J.~I.~B.~Camargo$^{1,2}$, % Observer
G.~Benedetti-Rossi$^{10,2}$, % Observer
M.~Kretlow$^{6}$,
R.~Vieira-Martins$^{1,2}$ % group leader
}% Solicitado por Josselin e Richard, participou da campanha
%P.~Roche$^{13}$ não encontrei o e-mail com a sugestão desse autor, 

%%%%%%%%% Insert author address here
\address{\scriptsize{
$^{1}$Observatório Nacional/MCTI, Rio de Janeiro, RJ, Brazil;
$^{2}$Laboratório Interinstitucional de e-Astronomia - LIneA, Rio de Janeiro, RJ, Brazil;
$^{3}$Federal University of Technology - Paraná (PPGFA/UTFPR), Curitiba, PR, Brazil;
$^{4}$LESIA, Observatoire de Paris, Université PSL, Sorbonne Université, Université de Paris, CNRS, 92190 Meudon, France;
$^{5}$Universidade Federal do Rio de Janeiro - Observatório do Valongo, Rio de Janeiro, RJ, Brazil;
$^{6}$Instituto de Astrofísica de Andalucía – Consejo Superior de Investigaciones Científicas, Granada, Spain;
$^{7}$British Astronomical Association, United Kingdom;
$^{8}$Institut Polytechnique des Sciences Avancées IPSA, 94200 Ivry-sur-Seine, France;
$^{9}$Institut de Mécanique Céleste et de Calcul des Éphémérides, IMCCE, Observatoire de Paris, PSL Research University, CNRS, Sorbonne Universités, UPMC Univ Paris 06, Univ. Lille, France;
%$^{9}$Internationale Amateursternwarte (IAS) e. V., Mittelstr. 6, 15749 Mittenwalde, Germany;
%$^{10}$International Occultation Timing Association - European Section (IOTA/ES), Am Brombeerhag 13, 30459 Hannover, Germany;
$^{10}$UNESP - Sao Paulo State University, Grupo de Dinamica Orbital e Planetologia, Guaratinguetá, SP, Brazil;
}}
%%%% Subject entries to be placed here %%%%
\subject{solar system, observational astronomy}

%%%% Keyword entries to be placed here %%%%
\keywords{stellar occultations, cometary coma, centaurs, comets}

%%%% Insert corresponding author and its email address}
\corres{Chrystian Luciano Pereira\\
\email{chrystianpereira@on.br}}

%%%% Abstract text to be placed here %%%%%%%%%%%%
\begin{abstract}
% 200 words
Comets offer valuable insights into the early Solar System's conditions and processes. Stellar occultations enables detailed study of cometary nuclei typically hidden by their coma. Observing the star's light passing through the coma helps infer dust's optical depth near the nucleus and determine dust opacity detection limits. 29P/Schwassmann-Wachmann 1, a Centaur with a diameter of approximately 60 km, lies in a region transitioning from Centaurs to Jupiter-Family comets. Our study presents the first-ever observed occultation by 29P, allowing in the future a more refined orbit and thus better predictions for other occultations. The light curve reveals a solid-body detection lasting $3.65\pm0.05$~seconds, corresponding to a chord length of approximately 54~km. This provides a lower limit for the object's radius, measured at $27.0\pm0.7$~km. We identified features on both sides of the main-body occultation around 1,700~km from the nucleus in the sky plane for which upper limits on apparent opacity and equivalent width were determined. Gradual dimming within 23~km of the nucleus during ingress only is interpreted as a localised dust cloud/jet above the surface, with an optical depth of approximately $\tau \sim 0.18$. 
\end{abstract}
%%%%%%%%%%%%%%%%%%%%%%%%%%%

%%%%%%%%%% Insert the texts which can accomdate on firstpage in the tag "fmtext" %%%%%
\maketitle
\section{Introduction}
\label{sec:Introduction}

Comets, as remnants from the early Solar System, provide invaluable insights into the conditions and processes that shaped our celestial neighbourhood. They are typically classified into two categories based on their orbital characteristics: long-period comets (LPC), which have highly elliptical orbits with long orbital period ($\mathrm{P} > 200$~years), typically originated from the distant Oort Cloud ($\sim10^4 - 10^5$~au) \cite{Nesvorny2018}. The short-period comets are defined as objects that have short orbital periods of less than $200$~years \cite{Nesvorny2018}. Examples include the Halley-type comets (HTC), with $20 < P < 200$~years and likely to be originated in the Oort Cloud \cite{Wang2014}, and the Jupiter-Family Comets (JFC), comets with $P < 20$~years that are dynamically affected by Jupiter and likely to originate from the Kuiper Belt region. 

As a Kuiper Belt Object (KBO) evolves inward to become a JFC, these bodies can be temporarily residing in chaotic orbits between Jupiter and Neptune, originating the Centaur class \citep{Horner2004}. The origin of Centaur objects is not definitively established, but it is hypothesized that they may evolve from Scattered Disk Objects (SDOs) or other Trans-Neptunian Object (TNO) populations \citep{VolkMalhotra2008}. This evolution likely begins with their migration into Neptune-crossing orbits, followed by gravitational scattering into the regions influenced by Jupiter and Saturn \citep{Morbidelli2008}, where at least one-third of the Centaurs will actually be injected into the JFC population \citep{Tiscareno2003}.

The object designated 29P/Schwassmann-Wachmann 1 (hereafter 29P) has captivated astronomers' attention due to its intriguing behaviour and dynamic nature. It was discovered in 1927 at Hamburg Observatory, Germany, by Arnold Schwassmann and Arno Arthur Wachmann while active and was then classified as a comet. It orbits the Sun in a quasi-circular orbit with a semi-major axis of $a = 5.98$~au, and low inclination $i = 9.4^{\circ}$, with an orbital period of approximately 14 years (JPL~K192/80). So, disregarding the frequent material ejections by the nucleus, we find the 29P in the Centaur classification. More precisely, this object lies in a region called ``Gateway'' - where the objects are transitioning from Centaurs to the JFC \cite{Sarid_2019}. 
In terms of size, \cite{Cruikshank1983} report a diameter of $40\pm5$~km for 29P, derived from observations using the NASA Infrared Telescope Facility (IRTF). Observations from the Spitzer Space Telescope indicate a diameter of $54 \pm 10$~km \citep{Stansberry2004}, while WISE measurements suggest a diameter of $46\pm13$~km \cite{Bauer2013}. Additionally, \cite{Schambeau2015} estimate a diameter of $D = 60.4^{+ 7.4}_{- 5.8}$~km, with a more refined value of $D = 64.6 \pm 6.2$~km \cite{Schambeau2021}, both obtained by modelling and removing the coma contribution. Stellar occultation observational campaigns were triggered in the United States by the RECON (Research and Education Collaborative Occultation Network) after the orbit of the 29P be updated using the December 05, 2022 observation reported in this paper. The 29P nucleus was detected in three events between December 19, 2022, and January 28, 2023, with four and five solid-body detections, respectively. The dimensions of the apparent ellipses fitted to the chords is $67.9 \times 53.3$~km on January 28, 2023, and $78.2 \times 48.2$~km, on December 19, 2022 \citep{Buie2023LPI}. This dimensions give a equivalent diameter of $60.2~\mathrm{km} < \mathrm{D_{equiv}} < 61.4~\mathrm{km}$.

%The estimated diameter of 29P varies according to different sources: 
%\cite{Cruikshank1983} provide a value of $40\pm5$~km \st{based on thermal measurements}; \cite{Stansberry2004} estimate it to be $54 \pm 10$~km; \cite{Bauer2013} from WISE measurements estimate it to be $46\pm13$~km. \cite{Schambeau2015} estimate a diameter of $D = 60.4^{+ 7.4}_{- 5.8}$~km. \textcolor{blue}{A more recent value of $D = 64.6 \pm 6.2$~km was obtained for the nucleus by modelling and removing the coma \cite{Schambeau2021}}. 

Stellar occultations, the passage of a Small Solar System Body in front of a star, offer a unique opportunity to study the cometary nuclei in detail, usually hidden beneath their coma. 
By observing the star's light as it passes through the cometary coma, astronomers can infer the optical depth of the dust close to the nucleus or determine detection limits for the dust opacity \citep{Combes1983}.
%valuable information about the comet's size, shape, and density, and even detect the presence of debris fields and dust clouds close to the nucleus 
As an active comet presents an extensive coma, the photocentre of the object can not be retrieved accurately. This implies low precision in the comet astrometric position and significant errors in the stellar occultation predictions, a problem that is overcome when occultation by the cometary nucleus is detected. In this sense, is more common to observe an appulse (or the occultation by the comet atmosphere) than a occultation by the nucleus of the comet. %The first example on this is the appulse by the comet 1P/Halley in 1909 \citep{Archenhold1910}. 
An appulse were observed by the comets Levy (1990 XX), with a impact parameter of $3,500$~km and a coma optical depth of $\tau = 0.4$ \citep{Rosenbush1994}. Aiming to search for absorption spectra within the coma, the atmospheres of comets 17P/Holmes and C/2007 W1 (Boatinni) were probed using appulses between November 2007 and September 2008 \citep{OMalia2010}. The 17P/Holmes coma was also detected using stellar occultation in October 2007, where the starlight dimming indicating an optical depth of $\tau = 0.04$ within $\sim 1,770$~km (1.5 arcseconds), with a thick dust coma within 0.01~arcseconds (12~km) \citep{Lacerda2012}.
In October 1996, a stellar occultation by the comet Hale–Bopp (C/1995 O1) reveals a dense coma with $\tau \geq 1$ within 20--70~km from nucleus center \citep{Fernandez1999}.

This work presents the results of the first-ever observed occultation by 29P, which results in determining the astrometric position with accuracy never achieved for this body. This allows for the refinement of the orbit in the future and more precise stellar occultation predictions to be made. We also searched for additional material in the object's vicinity, identifying some features at distances of about 1,700~km from the nucleus. The gradual dimming of the star during the ingress was interpreted as a dust cloud or a jet-like feature close to the nucleus extending at least 23~km above the surface, with an optical depth $\tau' = 0.18 \pm 0.02$. We also determine upper detection limits on apparent opacity and apparent equivalent width. 

%%%%%%%%%%%%%%% End of first page %%%%%%%%%%%%%%%%%%%%%

\section{Prediction and observation}
\label{sec:Prediction}

\begin{figure}
    \centering
    \includegraphics[width=0.9\hsize]{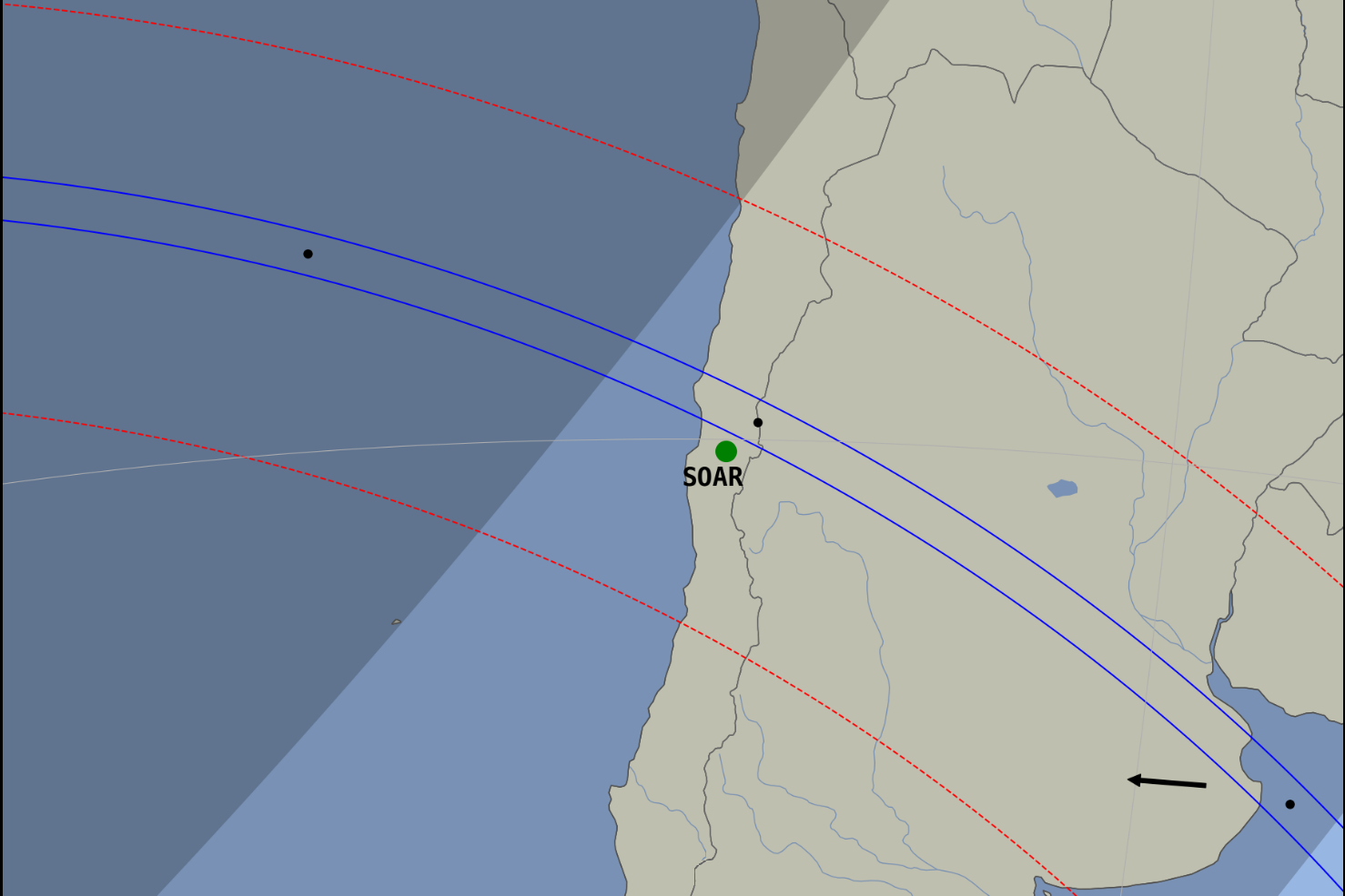}
    \caption{Prediction map with the shadow path considering the estimated radius (blue continuous lines). The black dots are separated by 60 seconds from each other. %, with the big dot representing the geocentric closest approach time.
    The dashed red line limits the $1\sigma$ uncertainty in the path. The arrow indicates the shadow direction of movement. The first version of this prediction was published on the \href{https://lesia.obspm.fr/lucky-star/occ.php?p=80809}{Lucky Star web page}. Note that the occultation occurred in twilight at SOAR. This complication was managed by adjusting the exposure time.}
    \label{fig:pred_map}
\end{figure}

The stellar occultation investigated in this work was predicted by the Lucky Star project using Numerical Integration of the Motion of an Asteroid (NIMAv2, based on 559 Earth-based direct observations) ephemeris\footnote{\url{https://lesia.obspm.fr/lucky-star/obj.php?p=997}} \citep{desmars2015} and Gaia Data Release 3 (Gaia DR3) sources \citep{GaiaColab2023}. The shadow path was predicted to cross the central regions of Chile and Argentina on December 05, 2022, at 08:14:19~UTC, as presented in Figure \ref{fig:pred_map}. The star (Gaia DR3 888441442906065536 source) has a magnitude G = 16.9. This source's Re-normalised Unit Weight Error (RUWE)\footnote{The Renormalised Unit Weight Error (RUWE) is a measure of the reliability of a single-star model derived from observations. A value close to 1 is typically expected. Values exceeding 1.4 could indicate that the source is not a single star or that there are problems with the astrometric solution \citep{LindegrenDR2}.} equals 1.027, and there is a false flag for duplicity in the catalogue, indicating that the star is not a double. At the time of the event, the apparent motion of the comet projected on the sky plane was 0.24"/min in position angle $274^\circ$, which equates to a geocentric velocity of $14.8~\mathrm{km\,s}^{-1}$, giving an expected maximum duration for the centrality of 4.1 seconds for a body diameter of 60 km \citep{Schambeau2015}. The star's equatorial coordinates were propagated to the event epoch using the proper motion and parallax \citep{GomesJr2022}, resulting in the geocentric position in the equinox of date:

\begin{center}
    ${\rm{RA}}: 6^h\,50^m\,43.88513^s \pm 0.49\,\mathrm{mas}$, \\
    ${\rm{DEC}}: +29^{\circ}\,23'\,47.8773'' \pm 0.49\,\mathrm{mas}$.  
\end{center}

Before the event, 29P’s nucleus underwent a series of notable outbursts, including a particularly strong one on 2022 November 21.95 $\pm$ 0.30, followed by two additional outbursts on 2022 November 27.69 $\pm$ 0.15 and 2022 November 29.15 $\pm$ 0.03, the latter occurring just 6.2 days before the predicted occultation (data from MISSION 29P\footnote{\url{https://britastro.org/section_information_/comet-section-overview/mission-29p-2/latest-lightcurve-plot-of-29p}}). Although the November 27 and November 29 outbursts were comparatively weaker, their proximity in time to the occultation increased the likelihood that residual ejecta remained in a temporary orbit close to the nucleus.

The observation occurred at the Southern Astrophysical Research Telescope (hereafter SOAR), located at Cerro Pachón, Chile. This telescope has a primary mirror with a diameter of 4.1-meter and Ritchey-Chrétien f/16 optics. The images were acquired in Flexible Image Transport System (FITS) format through an open filter wheel using the Raptor 247 Merlin camera, developed by Raptor Photonics, as the visitor instrument at the SOAR Observatory. This camera features a field of view of approximately $1.2' \times 0.9'$ and utilizes GPS for precise time synchronization. The data set contains 2,998 science images with 0.5 seconds of exposure time and about 100 calibration images to correct biases and pixel-to-pixel sensitivity variations. The dead time of the Raptor camera is negligible. 
%The observer also took some images of the comet separated from the target star right after the stellar occultation images using an exposure time of 2 seconds, with a negligible readout time. These images were co-added and enhanced to reveal possible jet-like features due to 29P outbursts.
%\begin{itemize}
%    \item Add table with observational circumstances?
%    \item Não, mas adicionar o mapa de predição e link.
%\end{itemize}

\section{Data Analysis and instants determination}
\label{sec:DataAnalysis}

The initial data set undergoes meticulous calibration through the classical correction by flat and bias frames using the Image Reduction and Analysis Facility \cite[IRAF,][]{Butcher1981}. This procedure ensures the reduction of artefacts and inaccuracies introduced during image acquisition. Photometric analysis is then conducted on these calibrated frames using the Package for the Reduction of Astronomical Images Automatically photometric task \citep[PRAIA,][]{Assafin2023}. %, with one star selected as the target. 
We detect contamination from the cometary coma during the aperture photometry performed on the target star. To effectively eliminate this contamination and obtain a light curve without significant interference from the scattered light of the coma entering the photometric aperture, we process each image in the data set individually.

\begin{figure*}[h]
    \centering
    \includegraphics[width=0.45\hsize]{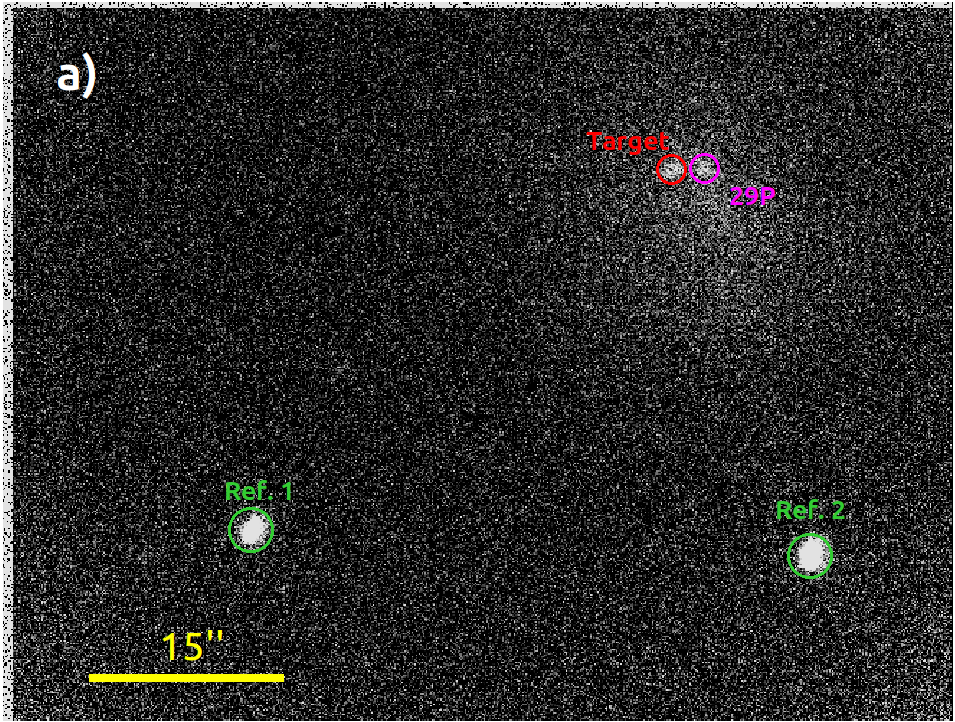}
    \includegraphics[width=0.45\hsize]{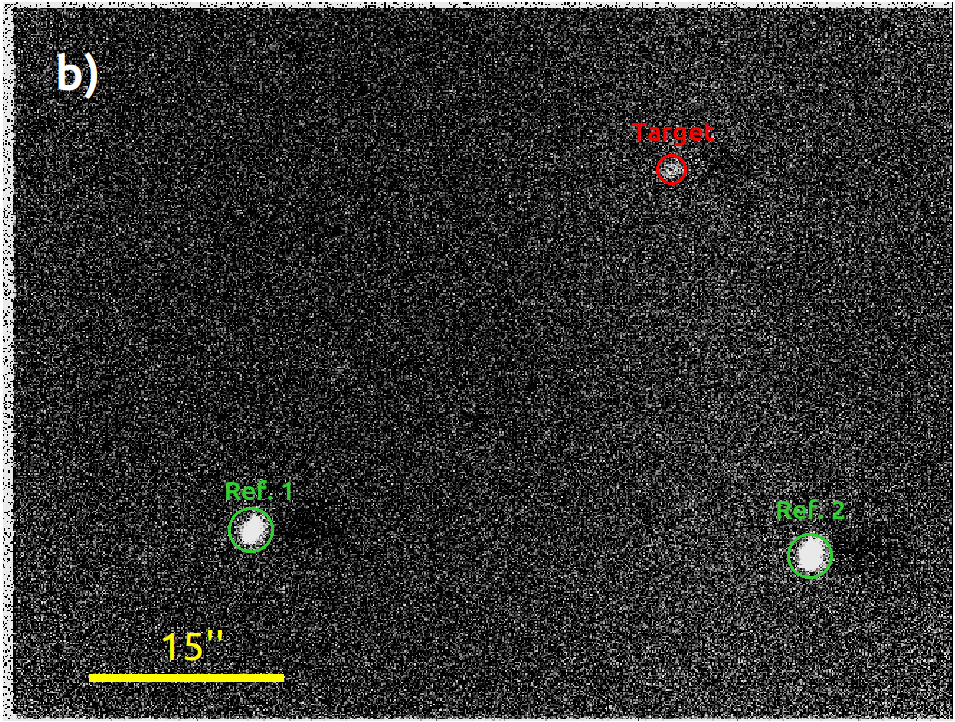}
    \includegraphics[width=0.45\hsize]{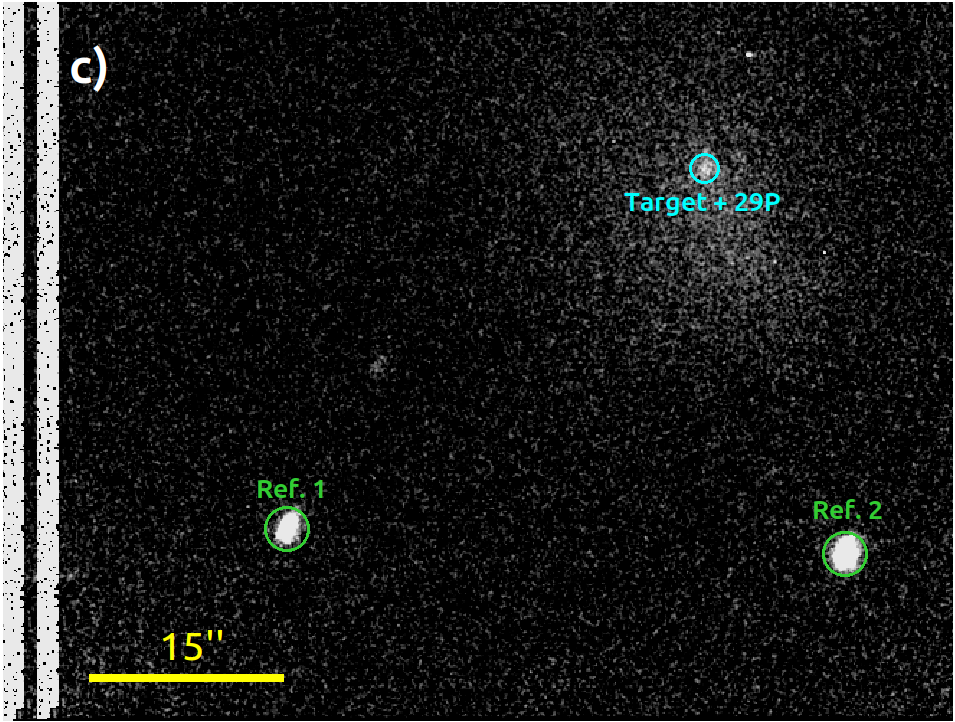}
    \includegraphics[width=0.45\hsize]{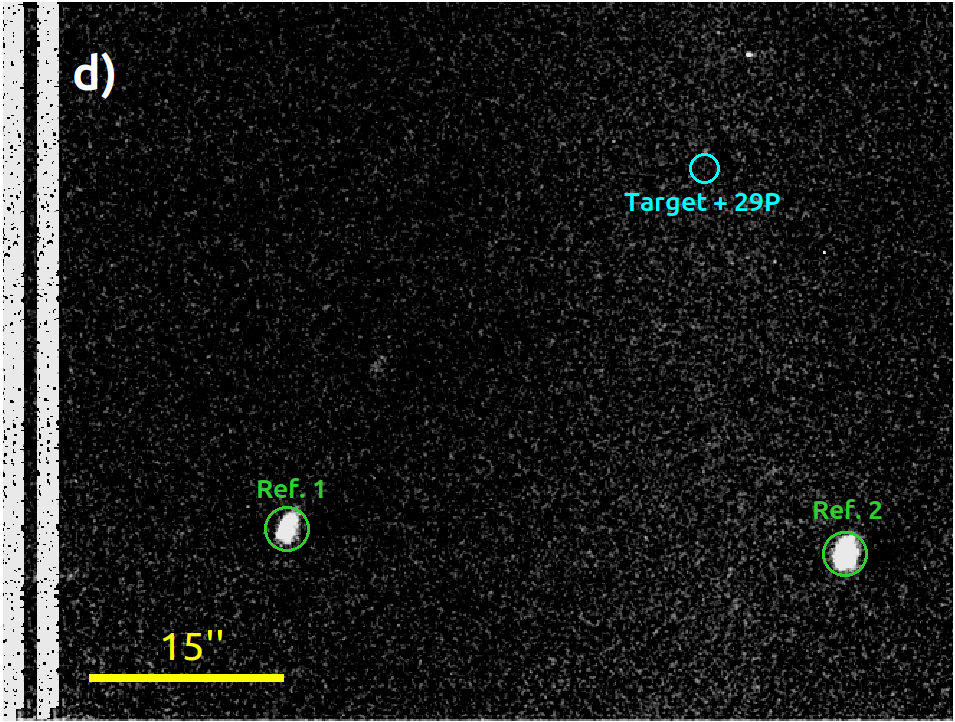}
    \caption{Elimination of the contamination from the cometary coma. First frame of the data set before (a) and after (b) the coma removal process. The frames c and d are in the central instant for the occultation, before and after the coma removal process, respectively. The red circle indicates the target star separated from the occulting body (magenta region). The cyan regions show the target star plus occulting body flux. The green regions indicate the reference stars used during the differential aperture photometry. Each frame has a field of view (FOV) of $\sim 1.2' \times 0.9'$.}
    \label{fig:comparison_frames}
\end{figure*}

The $x$ and $y$ positional data obtained from the photometric analysis are employed to align the stars within the images precisely. These aligned frames isolate the relative motion of the moving body, thereby facilitating a focused analysis of their behaviour against the fixed stellar background. %The next step involves conducting a fresh round of photometric analysis on the individual frames, this time with the comet as the target object. This yields accurate positional information of the comet within each image.
Assuming that the object’s trajectory appears as a straight line within the observed field, we can fit a first-degree polynomial to obtain theoretical positions in each frame, enhancing the precision of subsequent analysis. Using these theoretical positions, a composite frame is generated using median stacking in the comet's rest frame, ensuring that the stars disappear and providing a clear representation of the cometary coma. The last step is subtracting this stacked frame from each frame, as exemplified in Figure \ref{fig:comparison_frames}. %This effectively eliminates the contamination stemming from the comet’s coma, enabling the light curve of stellar occultation to be obtained without critical contamination from the scattered light of the comet entering the photometric aperture.

%Frames indicating the elimination of the contamination from the cometary coma in the first frame of the data set (top row) and the frame during the stellar occultation (bottom row). The left column presents the original frames, with the target star (red circle) separated from the 29P (magenta circle) and the star plus 29P flux (cyan circle). The right column presents the coma-corrected frames. The green circles indicate the comparison stars in the field used during the differential aperture photometry. \textbf{Each frame has a field of view (FOV) of $\sim 1.2' \times 0.9'$.}

\begin{figure}[ht]
    \centering
    \includegraphics[width=\hsize]{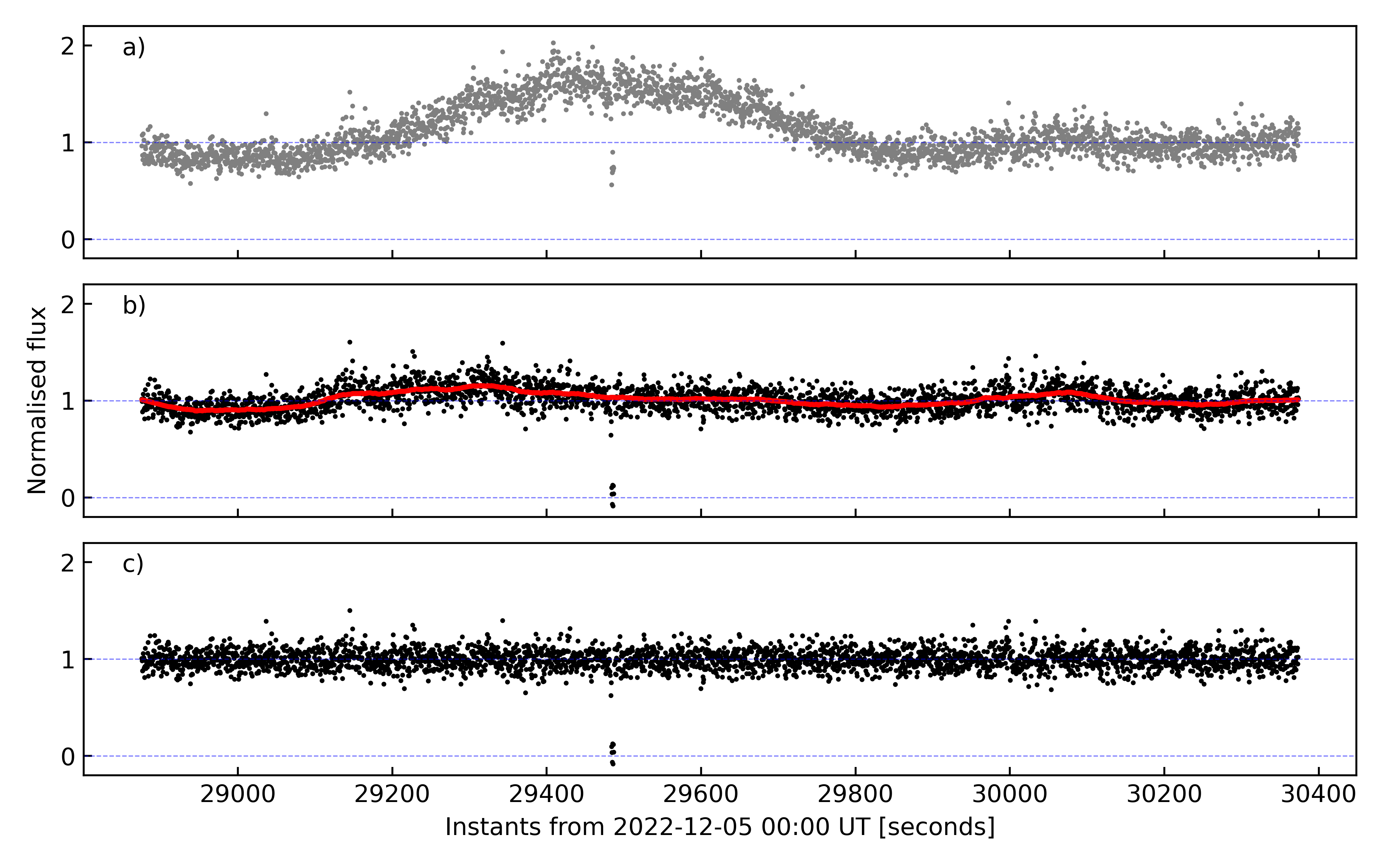}
    \caption{Light curves obtained from the aperture photometry. a) light curve obtained from images before the cometary coma is removed. b) Un-detrended light curve (black dots) obtained from differential photometry in the frames corrected from cometary coma plotted over the resampled light curve obtained with the Savitzky-Golay (SG) digital filter (red). c) Normalised light curve after the detrending process.}
    \label{fig:detrend_curve}
\end{figure}

Finally, we can perform the differential aperture photometry using the coma-eliminated frames. The target star flux was measured with a circular aperture, and the other two stars in the field were used as calibrators to mitigate the sky's fluctuations. 
%As photometry was conducted using images with the comet's coma removed, the light curve exhibits low-frequency systematic flux variations due to this correction. When stacking images centred on the comet using the median counts, the background stars were "blurred" in the field. Subtracting this median from the frames to be corrected reveals the formation of a shadow near the stars. Despite this shadow causing a variation in the flux of the target star and being partially corrected when considering the flux ratio of the target star and calibrators, the contamination effect is much smaller than that observed when conducting photometry on frames not corrected for the presence of the comet's coma. 
To normalise the light curve, we follow these steps: i) From photometry, we obtain the target and the calibrators' fluxes (ADU), both corrected from the sky background; ii) These fluxes were normalised by dividing them by their respective medians, thus not affected by the outliers. We call this light curve un-detrended; iii) Using the Savitzky-Golay (SG) digital filter, we calculated the trend of these systematic flux variations using windows of 135 seconds ($\sim$ 1,000 km, for an average spatial resolution of 7.4~km) in the outer regions of the occultation. Subsequently, we interpolated the region within the occultation using a polynomial function; iv) By dividing the un-detrended light curve by the SG resampled light curve, we obtained the detrended light curve, where the systematic low-frequency flux variations were removed. A similar approach was used to decrease the systematic variations in transit light curves, improving the detection of exoplanets \cite{Boufleur2018}. 
The results of the coma removal and normalisation processes are presented in Figure \ref{fig:detrend_curve}. The current methodology excludes diffuse and broader structures. This intentional decision corrects slow variations in the light curve, although it may not consider the presence of diffuse material around the body extinguishing the stellar flux.

\begin{figure}[h]
    \centering
    \includegraphics[width=\hsize]{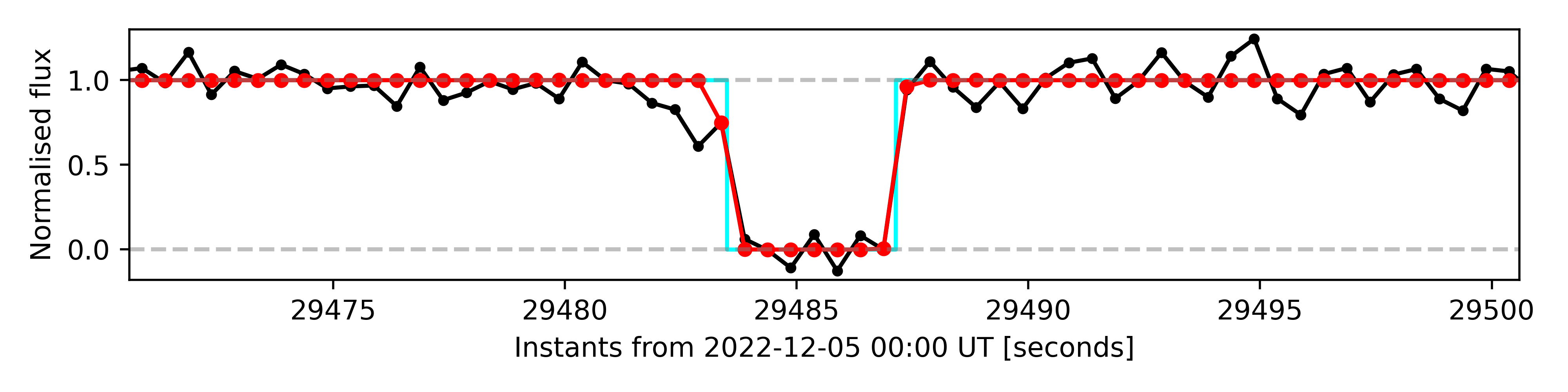}
    \caption{Modelled light curve (red) that best fits the observed data (black). The cyan curve represents the square-well model. Note the gradual drop in flux in the ingress region, explained in Section \ref{sec:GradualDimming}.}
    \label{fig:modeled_curve}
\end{figure}

We can now obtain the ingress and egress instants of the star behind the nucleus from the normalised light curve using the algorithms built using the Stellar Occultation Reduction and Analysis \citep[SORA,][]{GomesJr2022} package. 
A synthetic light curve is built considering an occultation by a sharp edge box convolved with the Fresnel diffraction, the apparent star diameter, observation wavelength, and the exposure time. 
We generate 100,000 models by performing a grid search over two parameters, which are varied to explore the solution space. These models are then fit to the observed light curve using chi-square ($\chi^2$) statistics to determine the best-fitting model. The fitting process includes N = 40 data points and two free parameters (M = 2), and the model with the minimal $\chi^2$ value is selected.
The ingress and egress instants are 08:11:23.00~$\pm$~0.06~UTC and 08:11:26.64~$\pm$~0.07~UTC , with an occultation duration of $3.65 \pm 0.09$~seconds and length of $54.2 \pm 1.3$~km. The minimum $\chi^2$ per degree of freedom is $\chi^{2}_{pdf} = 0.902$. The instrumental response dominates this light curve (0.5~seconds or 7.42~km). The calculated value for the Fresnel scale effect is 0.035~seconds or 0.52~km, and the star diameter effect is 0.013 km (for a star apparent diameter of 0.003~mas \citep{Kervella2004} projected at 29P's distance of 5.16~au). The best-fitted model is presented in Figure \ref{fig:modeled_curve}.

\section{Event geometry}
\label{sec:EventGeometry}

\begin{figure}[h]
    \centering
    \includegraphics[width=0.4\hsize]{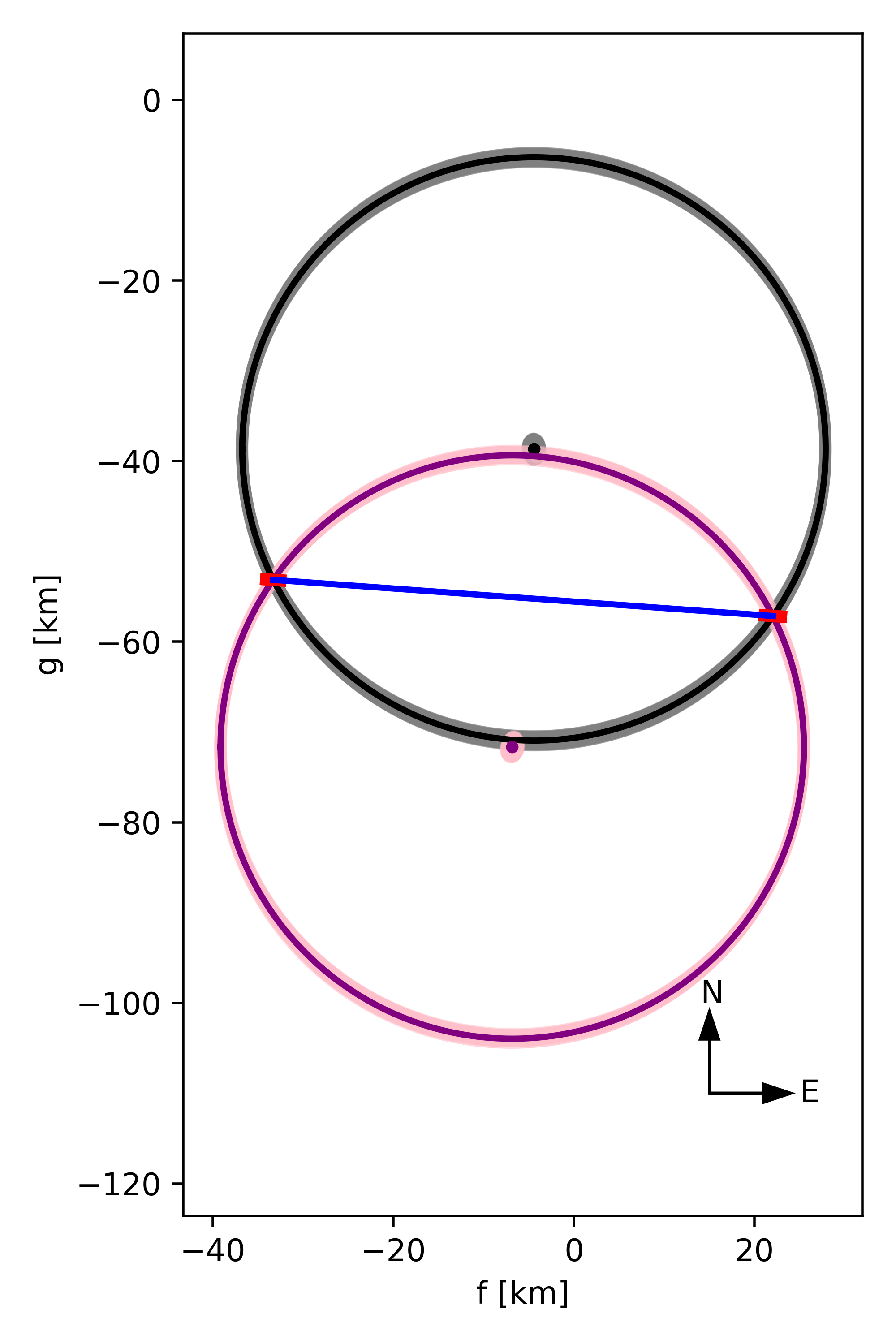}
    \caption{Two circles fitted to the SOAR chords showing the possible solutions for the 29P centre at event epoch. The south (resp. north) solution is in purple (resp. black), with the $1\sigma$ uncertainty in light purple (resp. grey) for the centre and the circular limb.}
    \label{fig:circles}
\end{figure}
Each chord extremity represents the intersection between the star position from an observer's point-of-view and the occulting body's silhouette. We can fit a circle to the chord extremities with only a single positive detection by varying the centre. For that we use the equivalent radius of ${\rm{R}_{equiv}} = 32.3$ obtained with Spitzer \citep{Schambeau2021}. 
%as a fixed value to Fitting a circle to one chord, keeping the radius fixed, gives us 
Two possible solutions for the circle's centre position ($f_\mathrm{c}, g_\mathrm{c}$) are possible: the north (resp. south) solution, with the centre located on the north (resp. south) side of the chord. In this context, the coordinates $f_0, g_0 = (0, 0)$  defines the ephemeris position of the target body in the closest approach. Thus, $f_\mathrm{c}, g_\mathrm{c}$ can be understood as an ephemeris offset. The best-fitted circles are presented in Figure \ref{fig:circles}. The purple circle (south solution) has centre position $f_\mathrm{c} = - 6.8\pm0.7$~km and $g_\mathrm{c} = -71.6\pm1.1$~km, and the black circle (north solution) has centre positions $f_\mathrm{c} = -4.4\pm0.7$~km and $g_\mathrm{c} = -38.7\pm1.1$~km. 
As north and south solutions are possible, both should be used as astrometric positions of 29P for orbit fit \cite{Rommel2020}, before further observations can show that one is preferable. Note that the difference between the two positions is only 6~mas; therefore, they represent a better position than those we can have from ground observation.
%Considering their uncertainties, we use the mean values between these two solutions for astrometric update purposes. 
The astrometric geocentric positions of 29P on 2022 December 05, at local closest approach time 08:11:24.8~UTC, are presented in Table \ref{tab:astrometric_position}.
%\begin{center}
%    North:\\
%    ${\rm{RA}}: 6^{h}\,50'\,43''.8769153 \pm 0.523\,mas,$ \\
%    ${\rm{DEC}}: + 29^{\circ}\,23'\,46''.530272 \pm 0.659\,mas.$  
%     \\South:\\
%      ${\rm{RA}}: 6^{h}\,50'\,43''.8768801 \pm 0.530\,mas,$ \\
%    ${\rm{DEC}}: + 29^{\circ}\,23'\,46''.523948 \pm 0.652\,mas.$  
%\end{center}
%
\begin{table}[ht]
    \centering
    \begin{tabular}{c c c} \hline \hline
    \multirow{2}{*}{Solution}   &   Right Ascension (h m s)    &   Astrometric uncertainty     \\ 
                                &   Declination  ($^\circ$ ’ ”)       &   ($\mathrm{mas}$)              \\ 
    \hline
    \multirow{2}{*}{North}  & 6 50 43.9298     & $\pm$0.5 \\
                            & +29 23 46.4766    & $\pm$0.6 \\
    \hline
    \multirow{2}{*}{South}  & 6 50 43.9299     & $\pm$0.5 \\
                            & +29 23 46.4678    & $\pm$0.6 \\
    \hline
    \end{tabular}
    \caption{Astrometric positions for the two solution of 29P centre for the geocentric closest approach epoch 2022 December 05 08:11:24.8~UTC.}
    \label{tab:astrometric_position}
\end{table}
\section{Gradual star dimming at ingress}
\label{sec:GradualDimming}
%We observe a gradual flux variation during the star's ingress behind the nucleus. 
We observe a gradual decline in flux during the ingress of the star behind the nucleus that appears absent at egress. This can be interpreted as 1) the star gradually occulted by the opaque body, revealing topographic features on the limb, or 2) a dense dust cloud close to the nucleus or a jet-like feature that partially blocks the stellar flux. These two hypotheses are addressed below.

The length of the SOAR chord is 54 km and considering the estimated equivalent diameter of 29P as $\mathrm{D_{equiv}}\sim64$ km, we can assume that the chord is nearly diametrical. If the gradual flux drop at the ingress is attributed to the star being partially occulted by 29P's limb, this would suggest a chord length of about 74 km. 
%For the star to be partially occulted, the chord would need to be tangential. In this case, a 74 km chord would imply an extremely elongated body, while still remaining consistent with the estimated equivalent diameter of around 62 km. 
However, the probability of a grazing event is very low given the chord length, and even more so considering the star’s size, which is nearly point-like in this case. 
%This interpretation is supported by the apparent elliptical dimensions from stellar occultations observed for this body a few days after the event reported in this work \cite[see][]{Buie2023LPI}. 
%Although unlikely, this topographical interpretation cannot be disregarded.
In this sense, the topographical interpretations seems to be unlikely.
The second option is analyse this feature as a semi-transparent screen representing a dust cloud over the surface or a jet-like structure.

\begin{figure}[h]
    \centering
    \includegraphics[width=\hsize]{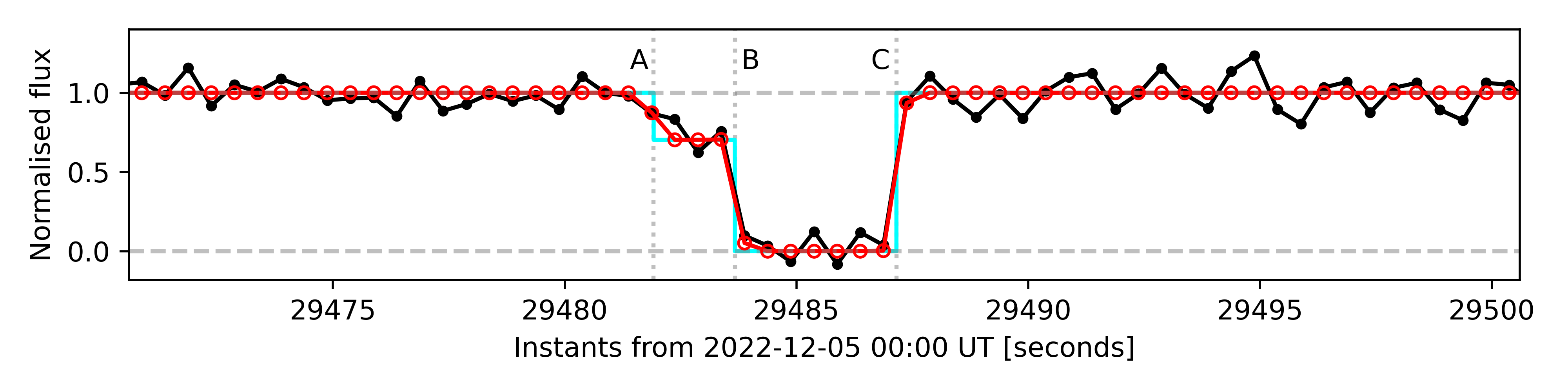}
    \caption{Modelled light curve (red) that best fits the observed data (black). The cyan curve represents the step-wise model for the semi-transparent screen blocking the stellar flux (segment A to B) right before the occultation by the nucleus (B to C).}
    \label{fig:two_box}
\end{figure}

This modelling is analogous to the procedure described in Section \ref{sec:DataAnalysis}, but besides the times of ingress and egress, we vary the box's opacity for the gradual dimming. The best synthetic light curve was obtained by $\chi^2$ statistics comparing the 100,000 generated models with the observed light curve, with the $1\sigma$ marginal error bar determined by the interval of $\chi^2_{min} + 1$. Figure \ref{fig:two_box} presents the step-wise model fitted to the proposed semi-transparent structure plus opaque nucleus.
The fitted semi-transparent box can be interpreted as a dust accumulation extending to a minimum height of $23.4$~km from the nucleus limb with an optical depth $\tau = 0.18 \pm 0.02$ (this optical depth takes into account the Airy diffraction \cite[see][]{Cuzzi1985}). Note that the occultation by the nucleus (segment BC) presents a slightly smaller chord length ($\sim$51.7~km) when compared with the previous fit, disregarding the gradual flux variation (Sec. \ref{sec:DataAnalysis}), which is not significant for the astrometric position.

\section{Detection limits on additional material}
\label{sec:DetectionLimits}

Equivalent width was employed to determine detection limits for additional material in light curves with arc detections \citep{Sicardy1991} during Neptune's stellar occultations (1983-1989). A similar approach was used to establish upper limits on apparent opacity $p'$ for structures around Pluto \cite{Boissel2014},  Chariklo \cite{berard2017, Morgado2021_Chariklo}, Quaoar \cite{Morgado2023, Pereira2023}, Chiron \citep{BragaRibas2023}, and Echeclus \citep{Pereira2024}. The equivalent width is defined as the width of an opaque strip that blocks the same amount of stellar light that a semi-transparent structure with width $W_r$. This quantity does not depend on the stellar apparent diameter or diffraction since both effects conserve energy \citep{Sicardy1991}. 

\begin{figure}[h]
    \centering
    \includegraphics[width=\hsize, trim={1cm 0 2cm 0},clip]{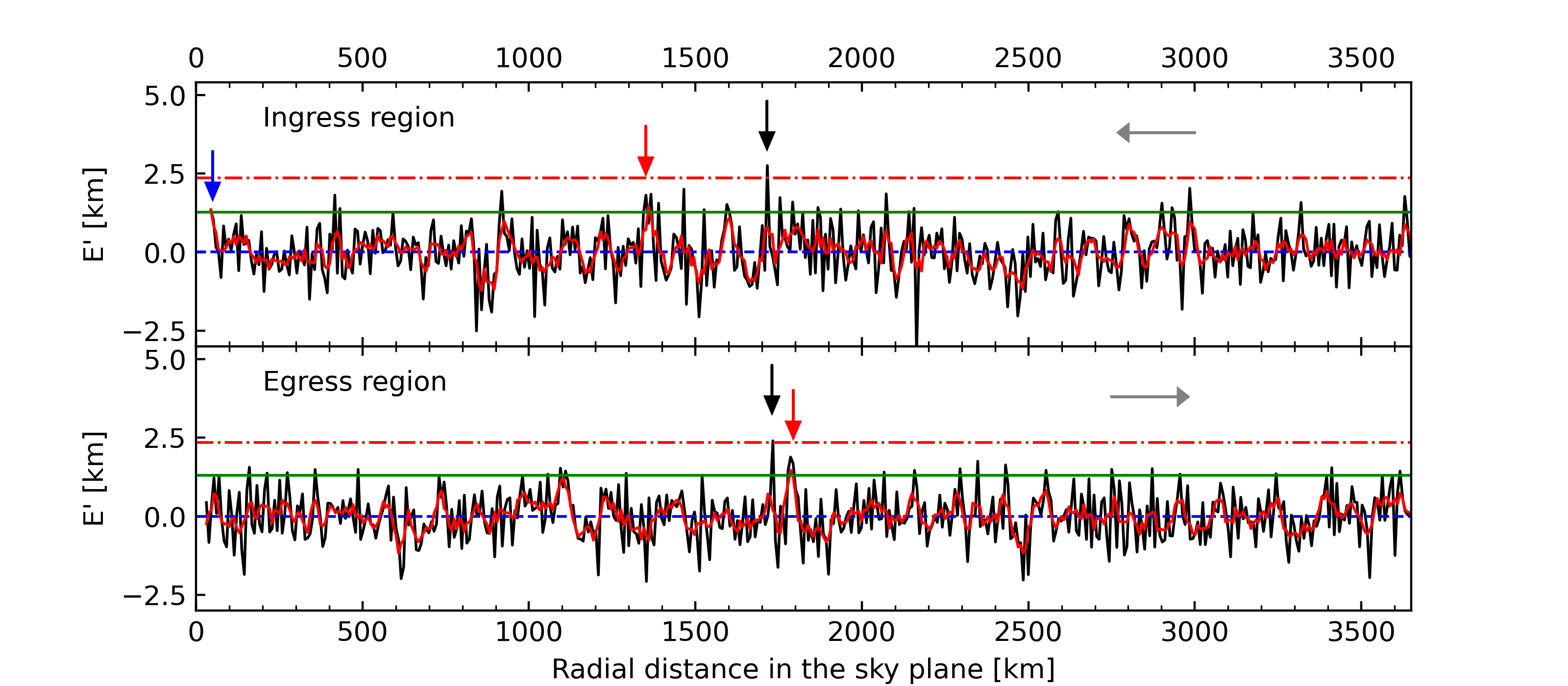}
    \caption{Apparent equivalent width as a function of the radial distance for the regions prior (top panel) and post (bottom panel) closest approach, covering a total of 7,000~km in the sky plane for better visualization. The grey horizontal arrows indicate the time evolution. The blue-dashed line indicates the mean of the ${\rm{E'}}$ distribution, and the red-dot-dashed line presents the $3\sigma$ upper limit of ${\rm{E'}}$. The green horizontal line indicates the $3\sigma$ upper limit for the 25~km-resampled curve (red). The arrows show the identified features over the upper limits for the full-resolution (black) and resampled (red) curves. We can also see the dust cloud/jet (Section 5) in the 25-km resampled curve (blue arrow).}
    \label{fig:E_limits}
\end{figure}

The upper limit on apparent equivalent width (apparent because we do not have a pole orientation for 29P, thus calculated on the sky's plane) is calculated by transforming the flux versus time light curve to apparent equivalent width ${\rm{E'}}(i)$ versus the radial distance in the sky plane using 
\begin{equation}
    {\rm{E'}}(i) = [1 - \phi(i)]\Delta r(i), 
    \label{eq:equiv_width_limits}
\end{equation}
where $\phi(i)$ is the normalised flux, $\Delta r(i)$ is the radial distance between consecutive points projected in the sky plane, and \textit{i} is the frame number. Then, the $3\sigma$ standard deviation is calculated for all the data points (N$_{\mathrm{pts}} = 2,988$) in the regions outside the main body occultation. The resulting curve is presented zoomed in Figure \ref{fig:E_limits}, depicting ${\rm{E'}}(i)$ against radial distance over about 7,000~km centred in 29P. Similarly, we calculated the $3\sigma$ standard deviation of the stellar flux $\phi(i)$ to determine the upper limits on apparent opacity $p'$. The $3\sigma$ upper limit on apparent equivalent width and apparent opacity is ${\rm{E'}}(i) \sim$ 2.3~km and $p' \sim$ 0.3, covering a radial distance in the sky plane of $\sim22,700$~km (or 6.0 arc-seconds). This implies that any structure with intermediate solutions between these two extreme cases would be detectable: an opaque structure with a width of about 2.3~km or a semi-transparent structure with a width of $W_{\perp} \sim\,$7.4~km (light curve mean spatial resolution) and with an apparent opacity $p' 0.3$.  

We identified features above to the $3\sigma$ cut at an average distance of 1,723~km from the object's centre, assuming that this centre is the mean value between the two possible circular solutions. At the ingress region (before the main event), the feature has ${\rm{E'}} = 2.8 $~km ($3.6\sigma$) at 1,715~km in the sky plane. The feature in the egress (after the main event) is located 1,730~km in the sky plane, with ${\rm{E'}} = 2.4$~km ($3\sigma$). This is equivalent to a semi-transparent structure with a width of $\sim7.6$~km and apparent opacities of $p'=0.368$ and $p'=0.316$, for ingress and egress, respectively. This distance is close to the co-rotational region ($\sim 1,757$~km), obtained considering the nucleus rotational period of 57.7~days (estimated from the periodicity of the outbursts \cite{MILES2016}), radius of $32.3$~km, and density $\rho=1,119\,{\rm{kg\,m}^{-3}}$ (based on estimates from active Centaur Chiron \citep{BragaRibas2023}). 
To assess the significance of the observed outliers, we performed a calculation assuming a Gaussian distribution for the data. For the $2,988$ data points, we would expect for about 9 data points beyond $3\sigma$ and about 1 data point beyond $3.5\sigma$ purely by random chance. While detections at the $3\sigma$ and $3.6\sigma$ levels are noteworthy, they do not independently indicate statistical significance. However, the symmetrical location of these outliers is intriguing and suggests a potential non-random pattern that warrants further investigation.

Using 25-km bins in our search for slightly larger structures, we found beyond our $3\sigma$ cut from the mean of the distribution, the semi-transparent feature close to the nucleus presented in Section \ref{sec:GradualDimming} and two other features.
%Using 25~km bins in our search for slightly larger structures, we found two features that stand out beyond our $3\sigma$ cut from the mean of the distribution. 
Unlike the structures previously presented, these have more points inside the flux drop, which increases the significance of these detections, standing at $3.8\sigma$ and $3.9\sigma$ of the mean for ingress and egress regions, respectively. We applied a square-box fit with variable apparent opacity to determine the properties of these potential detections. Located at $1,352.3\pm1.0$~km (resp. $1,796.8 \pm 2.2$~km) from the centre in the sky plane at ingress (resp. egress), these features have an average width of $W_\perp =29.4\pm2.2$~km and optical depth $\tau = 0.11\pm0.03$ (resp. $W_\perp =28.0\pm5.0$~km and optical depth $\tau = 0.13\pm0.06$), considering the individual diffraction of particles \cite{Cuzzi1985} and errors within $1\sigma$ level.

\section{Conclusion}
\label{sec:Conclusion}

This work presents the results of the first-ever observed stellar occultation by the Centaur 29P, observed with the SOAR telescope on December 5, 2022. The event involved a star with magnitude G = 16.9 while 29P was active and showed an extensive cometary coma. The acquired images were corrected to reduce the coma contamination, enabling the stellar occultation light curve to be extracted through differential aperture photometry. 

This single-chord event allows for a better determination of the 29P's astrometric position, improving its orbit and predictions for future stellar occultations. Our analysis indicated that the positional error has been significantly reduced from roughly 1,600~km (0.375~arcsec)\footnote{\url{https://lesia.obspm.fr/lucky-star/obj.php?p=997}} to about 260~km (0.061~arcsec)\footnote{\url{https://lesia.obspm.fr/lucky-star/obj.php?p=1009}} after a 1-year orbital arc. Notably, this increased precision could have broader implications for understanding the long-term evolution of the object, as 29P resides in a dynamically chaotic region of phase space known as the JFC Gateway region, as defined by \cite{Sarid_2019}. The chaotic nature of this region suggests that even small changes in orbital parameters could impact the body's secular evolution studies. This opens the possibility for future work to explore whether 29P might transit between dynamical classes or remain within stable pockets (lobes) of phase space \cite[see][]{Swenson2019}. This aspect, however, is beyond the scope of the current work.

The nucleus of 29P exhibited three significant outbursts between 6 to 13 days prior to the occultation: the first on 2022 November 21.95 $\pm$ 0.30, which had an amplitude of 3.75 mag, followed by an outburst on 2022 November 27.69 $\pm$ 0.15 with an amplitude of 1.65 mag, and a smaller outburst on 2022 November 29.15 $\pm$ 0.03, with an amplitude of 0.32 mag.
Specifically, the intensities for these three were 164, 96, and 32 nucleus-equivalents, respectively, where one nucleus-equivalent is the absolute magnitude R(1,1,0) of the nucleus taken to equal $9.77 \pm 0.04$. This value is based on unpublished coma profile observations of 29P (38 epochs of observation obtained during quiescence using 2.0-m telescopes between 2014 and 2023).

Considering the densities range $400$--$1,118~\mathrm{kg\,m^{-3}}$ \citep{Britt2006,BragaRibas2023} and the equivalent radius of $\mathrm{R_{equiv}} = 32.3$~km \citep{Schambeau2021}, the expected velocity of ejected material measured normal to the surface of 29P is $15$--$25\,\mathrm{m\,s^{-1}}$. Models indicates that the dust ejection velocity from 29P's surface, at a heliocentric distance of $\sim6$~au, lies between $10$--$20\,\mathrm{m\,s^{-1}}$ \cite{Fulle1992}. Measurements on particles ejected by the comet 1P/Halley at $\sim14$~au reveals a velocity of $14.5\,\mathrm{m\,s^{-1}}$ \cite{West1991}. Studies on 174P/Echeclus direct images obtained in 2016\footnote{\url{https://britastro.org/cometobs/174p/174p_20160905_rmiles.html}} results in a estimated velocity of about $95\,\mathrm{m\,s^{-1}}$. From an outburst in the comet 67P/Churyumov-Gerasimenko observed on 2016 July 03 with several instruments onboard ESA's Rosetta spacecraft, the dust particles velocity were calculated in the range $0.41$--$25\,\mathrm{m\,s^{-1}}$ \cite{Agarwal2017}.
%For comparison, features observed near to the 29P nucleus during the Hubble Space Telescope observations in March 1996 corresponds to an average expansion velocity of about $0.017\,\mathrm{km\,s^{-1}}$ \citep{MILES2016b}. 
This implies that material expelled during outbursts may result in distinct outcomes for the ejected particles, with some returning to the nucleus while others potentially achieving escape velocity and being ejected.
%
%\st{Therefore, we attribute the dimming to a localised emission of dust and gas from an area on the nucleus near the morning terminator, solar radiation heating that region and provoking the activity.} 
%\textcolor{blue}{If we assume that the activity generating this feature occurs on the morning side of the nucleus, it would imply a retrograde spin. However, we primarily attribute the dimming to localized dust cloud related to solar heating.} In this case, the observation would suggest the nucleus spins in a retrograde direction. This behaviour was interpreted as a potential dense dust cloud near the nucleus, extending for a height $> 23$~km above the surface and with an optical depth of $\tau = 0.18 \pm 0.02$. 
If we assume that the activity generating this feature occurs on the morning side of the nucleus, and is thus an insolation-driven outflow, this would imply a retrograde spin. Recent observations of 29P with the James Webb Space Telescope (JWST) have indeed revealed localized jet features in three active regions, with the frontal jet associated with an outflow driven by solar heating \cite{Faggi2024}. With this scenario, we might be probing the densest part of this cometary jet in our light curve, either during the dust-lifting outgassing phase or through the accumulation of material (dust cloud) falling back onto the nucleus. Our analysis revealed a gradual dimming of the star during ingress, suggesting that the dust reaches a minimum height of about 23~km above the surface, with an optical depth of $\tau=0.18\pm0.02$. 

%Although less likely due to the small apparent size of the occulted star, there is still the possibility that this gradual variation in flux is due to the roughness of the limb (topography). 

%\st{When analysing the full-resolution light curve, we identify single data point flux drops that correspond to possible structures with apparent equivalent width ${\rm{E'}} = 2.8$~km and ${\rm{E'}} = 2.4$~km, for the regions prior and after the closest approach, respectively, and localised at an average distance of 1,723~km from the cometary nucleus in the sky plane. The 25-km resampled light curve shows flux variations over the $3\sigma$ upper limit: at $1,352.3\pm1.0$~km from the nucleus centre in the prior-closest approach region, the feature has a width $W_{\perp} = 29.4 \pm 2.2$~km and opacity $p = 0.10 \pm 0.03$. After the closest approach, the feature has a width $W_{\perp} = 28.0 \pm 5.0$~km and opacity $p=0.12 \pm 0.05$ . Despite the low significance of these features, the symmetry concerning the body's centre is noteworthy. While unlikely, these features might suggest a localised accumulation of material ejected by the nucleus, forming a debris envelope or other confined narrow structures.}

Flux drops characteristic of additional material were identified in the light curve, and despite their low significance, the symmetry concerning the body's centre is noteworthy. While unlikely, these features might suggest a localised accumulation of material ejected by the nucleus, forming a debris envelope or other confined narrow structures. When analysing the full-resolution light curve, we identify single data point flux drops that correspond to possible structures with apparent equivalent widths of ${\rm{E'}} = 2.8$~km and ${\rm{E'}} = 2.4$~km, distant about $1,723$~km from the cometary nucleus in the sky plane. In the 25-km resampled light curve, flux variations above the $3\sigma$ upper limit appear at distinct distances: at $1,352.3\pm1.0$~km from the nucleus centre in the prior-closest approach region, the feature has a width of $W_{\perp} = 29.4 \pm 2.2$~km and an optical depth $\tau = 0.11\pm0.03$. Following the closest approach, another feature is identified at $1,796.8\pm2.2$~km with a width of $W_{\perp} = 28.0 \pm 5.0$~km and an optical depth $\tau = 0.13\pm0.06$.
%
%\textcolor{blue}{We can contextualize the optical depth values obtained in this work with values derived for structures around other objects. Observations of an appulse of comet 17P/Holmes allowed for the optical depth of the dust coma to be constrained to $\tau = 0.003$--$0.004$ at $\sim1,770~\mathrm{km}$ (1.5 arcsec) and being optically thick at about $\sim12$~km over comet surface \cite{Lacerda2012}. In situ observations of the comet 67P/Churyumov-Gerasimenko using the OSIRIS (Optical, Spectroscopic and Infrared Remote Imaging System) instrument onboard Rosetta spacecraft were used to determine the optical depth of the plume in the May 2016 outburst, equal to $\tau \sim 0.65$ \cite{Fornasier2019}. Chiron may be the most observed active Centaur object using the technique of occultations. During an appulse in March 1994, the light curve revealed flux variations caused by additional material, which were interpreted as cometary jets exhibiting different optical depths ($\tau < 1$) \cite{Elliot1995}. Notably, the structure labelled F2 in Figure 1 of \cite{Elliot1995} shows a comparable optical depth of $\tau \sim 0.11$. 
%Overall, our findings contribute to a deeper understanding of 29P/Schwassmann-Wachmann 1 and underscore the importance of stellar occultations in probing the intricate nature of small solar system bodies. Future studies building upon these results could further elucidate the dynamics and evolution of comets in our solar system.
The optical depth values of the structures proposed in this work can be contextualized within the range of values reported for dust structures around other small bodies. For example, observations of an appulse of comet 17P/Holmes constrained the optical depth of its dust coma to approximately $\tau = 0.003$--$0.004$ at a distance of $\sim1,770$~km (1.5 arcsec), with regions near the comet surface ($\sim~12$~km) becoming optically thick \cite{Lacerda2012}. In situ data from the OSIRIS instrument on board the Rosetta spacecraft enabled a measurement of optical depth of a plume during a May 2016 outburst on comet 67P/Churyumov-Gerasimenko, with an estimated value of $\tau \sim 0.65$ \cite{Fornasier2019}. Centaurs such as Chiron, frequently observed through stellar occultations, offer further insights. For instance, an appulse observed in March 1994 revealed light curve variations attributed to material interpreted as cometary jets, with optical depths below unity ($\tau < 1$) \cite{Elliot1995}. However, subsequent observations suggest the presence of a ring system around Chiron \cite{Ruprecht2015,Ortiz2015,Sickafoose2020,Sickafoose2023,Ortiz2023}. Notably, the F2 feature in Figure 1 of \cite{Elliot1995} shows an optical depth close to $\tau \sim 0.11$, comparable to some of our findings. 

Our study contributes to the growing understanding of 29P/Schwassmann-Wachmann 1, demonstrating the value of stellar occultations in revealing the complex behaviours and structures of small bodies. Future research building on these observations can help clarify the dynamic processes and evolutionary pathways of comets in the Solar System.

\vskip6pt

\dataccess{
The data is available at CDS via anonymous ftp to cdsarc.u-strasbg.fr
(130.79.128.5) or via \url{https://cdsarc.cds.unistra.fr/viz-bin/cat/J/RSPTA}
}
\ack{
We are thankful to the two anonymous referees for their thorough review and thoughtful recommendations, which have helped us improve this paper.
C.L.P is thankful for the support of the Coordenaç\~ao de
Aperfeiçoamento de Pessoal de N\'ivel Superior - Brasil (CAPES) and FAPERJ/DSC-10 (E26/204.141/2022).
This work was carried out within the “Lucky Star” umbrella, which agglomerates the efforts of the Paris, Granada, and Rio teams. The European Research Council funded it under the European Community’s H2020 (ERC Grant Agreement No. 669416). 
This study was partly financed by the National Institute of Science and Technology of the e-Universe project (INCT do e-Universo, CNPq grant 465376/2014-2) and partly by CAPES – Finance Code 001. 
The authors acknowledge the respective CNPq grants: B.E.M. 150612/2020-6; F.B.R. 316604/2023-2; M.A. 427700/2018-3, 310683/2017-3, 473002/2013-2; J.I.B.C. acknowledges grants 305917/2019-6, 306691/2022-1 (CNPq) and 201.681/2019 (FAPERJ). 
J.L.O. and M.K. acknowledge financial support from the grant CEX2021-001131-S funded by MCIN/AEI/10.13039/501100011033 and the financial support from the Spanish grant nos. AYA-2017-84637-R and PID2020-112789GB-I00 and the Proyectos de Excelencia de la Junta de Andalucía grant nos. 2012-FQM1776 and PY20-01309.
Based on observations obtained at the Southern Astrophysical Research (SOAR) telescope, which is a joint project of the Ministério da Ciência, Tecnologia e Inovações do Brasil (MCTI/LNA), the US National Science Foundation’s NOIRLab, the University of North Carolina at Chapel Hill (UNC), and Michigan State University (MSU).}

%%%%%%%%%% Insert bibliography here %%%%%%%%%%%%%%

\bibliographystyle{RS}
\bibliography{biblio}

\end{document}